\documentclass[10pt]{article}
\usepackage{graphicx}
\usepackage{amsmath}
\usepackage{amssymb}
\usepackage{caption2}
\begin{document}
\begin{center}
{\large {\bf \sc{  Analysis of  the strong  decay $X(5568) \to B_s^0\pi^+$  with QCD sum rules }}} \\[2mm]
Zhi-Gang Wang \footnote{E-mail: zgwang@aliyun.com.  }     \\
 Department of Physics, North China Electric Power University, Baoding 071003, P. R. China
\end{center}

\begin{abstract}
In this article, we take the $X(5568)$  to be the scalar  diquark-antidiquark type tetraquark state,  study the hadronic coupling constant $g_{XB_s\pi}$ with  the three-point QCD sum rules by carrying  out the operator product expansion up to the vacuum condensates of dimension-6 and including both the connected and disconnected Feynman diagrams, then calculate the partial decay width of the strong decay  $ X(5568) \to B_s^0 \pi^+$ and obtain the value $\Gamma_X=\left(20.5\pm8.1\right)\,\rm{MeV}$, which is   consistent with the experimental data $\Gamma_X = \left(21.9 \pm 6.4 {}^{+5.0}_{-2.5}\right)\,\rm{MeV}$ from  the D0 collaboration.

\end{abstract}



\section{Introduction}

Recently,  the  D0 collaboration  observed   a narrow structure, $X(5568)$, in the decay  $X(5568) \to B_s^0 \pi^{\pm}$  with significance  of $5.1\sigma$ \cite{X5568-exp}.   The measured mass and  width  are   $m_X =\left(
5567.8 \pm 2.9 { }^{+0.9}_{-1.9}\right)\,\rm{MeV} $   and $\Gamma_X = \left(21.9 \pm 6.4 {}^{+5.0}_{-2.5}\right)\,\rm{MeV}$,  respectively.
The D0 collaboration fitted the $B_s^0 \pi^{\pm}$ systems with the Breit-Wigner parameters in relative S-wave,  the favored  quantum numbers are $J^P =0^+$. However,
the quantum numbers  $J^P = 1^+$ cannot be excluded according to  decays  $X(5568) \to B_s^*\pi^+ \to B_s^0 \pi^+ \gamma$, where the low-energy photon is not
detected.
There have been several possible assignments, such as the scalar-diquark-scalar-antidiquark type tetraquark state \cite{WangX5568,ChenZhuX5568,NielsenX5568-mass,WangZhuR-X5568,LiuZhuX5568,NielsenX5568-decay}, axialvector-diquark-axialvector-antidiquark type tetraquark state \cite{ChenZhuX5568,AziziX5568-mass,AziziX5568-decay}, $B^{(\ast)} \bar{K}$ hadronic molecule state \cite{Chen-BK-5568},  threshold effect \cite{ThresholdX5568}.

The calculations based on the QCD sum rules indicate that  both the  scalar-diquark-scalar-antidiquark type and axialvector-diquark-axialvector-antidiquark type interpolating currents can give satisfactory mass $m_X$ to reproduce the experimental data \cite{WangX5568,ChenZhuX5568,NielsenX5568-mass,AziziX5568-mass}.
In Ref.\cite{AziziX5568-decay},  Agaev, Azizi and Sundu choose the axialvector-diquark-axialvector-antidiquark type interpolating current, calculate the  hadronic coupling constant $g_{XB_s\pi}$ with  the light-cone QCD sum rules in conjunction with the soft-$\pi$ approximation  and other approximations, and obtain the partial  decay width for the process   $ X(5568) \to B_s^0 \pi^+$. In Ref.\cite{NielsenX5568-decay},   Dias et al choose the scalar-diquark-scalar-antidiquark type interpolating current, calculate the  hadronic coupling constant $g_{XB_s\pi}$ with  the three-point QCD sum rules in  the soft-$\pi$ limit by taking into account only the connected Feynman diagrams in the leading order approximation, and obtain the partial  decay width for the decay   $ X(5568) \to B_s^0 \pi^+$. In previous work \cite{WangX5568}, we choose the scalar-diquark-scalar-antidiquark type interpolating current to study the mass of the $X(5568)$ with the QCD sum rules. In this article, we extend our previous work to study   the hadronic coupling constant $g_{XB_s\pi}$ with  the three-point QCD sum rules  by carrying  out the operator product expansion up to the vacuum condensates of dimension-6 and including both the connected and disconnected Feynman diagrams, then calculate the partial decay width of the strong decay  $ X(5568) \to B_s^0 \pi^+$.

 The article is arranged as follows:  we derive the QCD sum rule for the  hadronic coupling constant $ g_{XB_s\pi}$ in Sect.2;  in Sect.3, we present the numerical results and discussions; and Sect.4 is reserved for our
conclusion.

\section{QCD sum rule for  the hadronic coupling constant $ g_{XB_s\pi}$ }

We can study the strong decay $X(5568)\to B_s^0\pi^{+}$  with the three-point correlation function
$\Pi(p,q)$,
\begin{eqnarray}
\Pi(p,q)&=&i^2\int d^4xd^4y e^{ip \cdot x}e^{iq \cdot y}\langle 0|T\left\{J_{B_s}(x)J_{\pi}(y)J_{X}(0)\right\}|0\rangle\, ,
\end{eqnarray}
where the currents
\begin{eqnarray}
J_{B_s}(x)&=&\bar{s}(x)i\gamma_5 b(x) \, ,\nonumber \\
J_{\pi}(y)&=&\bar{u}(y)i\gamma_5 d(y) \, , \nonumber \\
J_X(0)&=&\epsilon^{ijk}\epsilon^{imn}u^j(0)C\gamma_5 s^k(0) \bar{d}^m(0)\gamma_5 C \bar{b}^n(0) \, ,
\end{eqnarray}
interpolate the mesons $B_s$, $\pi$ and $X(5568)$, respectively,  the $i$, $j$, $k$, $m$, $n$ are color indexes, the $C$ is the charge conjugation matrix.
In Ref.\cite{NielsenX5568-decay}, the axialvector current is used to interpolate the $\pi$ meson.

At the hadron side, we insert  a complete set of intermediate hadronic states with
the same quantum numbers as the current operators
    $J_{B_s}(x)$,   $J_{\pi}(y)$ and $J_X(0)$ into the three-point
correlation function $\Pi(p,q)$ and  isolate the ground state contributions to obtain the following result,
\begin{eqnarray}
\Pi(p,q)&=& \frac{f_{\pi}m_{\pi}^2f_{B_s}m^2_{B_s}\lambda_{X}g_{XB_s\pi}}{(m_u+m_d)(m_b+m_s)} \frac{1}{\left(m_{X}^2-p^{\prime2}\right)\left(m_{B_s}^2-p^2\right)\left(m_{\pi}^2-q^2\right)}   \nonumber\\
&&+ \frac{1}{\left(m_{X}^2-p^{\prime2}\right)\left(m_{B_s}^2-p^2\right)} \int_{s^0_\pi}^\infty dt\frac{\rho_{X\pi}(p^2,t,p^{\prime 2})}{t-q^2}\nonumber\\
&&+ \frac{1}{\left(m_{X}^2-p^{\prime2}\right)\left(m_{\pi}^2-q^2\right)} \int_{s^0_{B_s}}^\infty dt\frac{\rho_{XB_s}(t,q^2,p^{\prime 2})}{t-p^2}+\cdots \, ,
\end{eqnarray}
where $p^\prime=p+q$, the $f_{B_s}$, $f_{\pi}$  and $\lambda_{X}$ are the decay constants of the mesons  $B_s$, $\pi$  and $X(5568)$, respectively, the $g_{XB_s\pi}$ is the hadronic coupling constant.

In the following, we write down the definitions,
\begin{eqnarray}
\langle0|J_{X}(0)|X(p^\prime)\rangle&=&\lambda_{X} \,\, , \nonumber \\
\langle0|J_{B_s}(0)|B_s(p)\rangle&=&\frac{f_{B_s}m_{B_s}^2}{m_b+m_s} \,\, , \nonumber \\
\langle0|J_{\pi}(0)|\pi(q)\rangle&=&\frac{f_{\pi}m_{\pi}^2}{m_u+m_d} \,\, ,  \\
\langle B_s(p)\pi(q)|X(p^{\prime})\rangle&=&ig_{X B_s\pi} \, .
\end{eqnarray}

The two unknown functions $\rho_{X\pi}(p^2,t,p^{\prime 2})$ and $\rho_{XB_s}(t,q^2,p^{\prime 2})$ have complex dependence on the transitions
between the ground state $X(5568)$ and the excited states  of the $\pi$  and $B_s$ mesons, respectively. We introduce the parameters $C_{X\pi}$  and $C_{X B_s}$ to parameterize the net effects,
\begin{eqnarray}
C_{X\pi}&=&\int_{s^0_\pi}^\infty dt\frac{ \rho_{X\pi}(p^2,t,p^{\prime 2})}{t-q^2}\, ,\nonumber\\
C_{XB_s}&=&\int_{s^0_{B_s}}^\infty dt\frac{\rho_{XB_s}(t,q^2,p^{\prime 2})}{t-p^2}\,  ,
\end{eqnarray}
and rewrite the correlation function $\Pi(p,q)$ into the following form,
\begin{eqnarray}
\Pi(p,q)&=&\frac{f_{\pi}m_{\pi}^2f_{B_s}m^2_{B_s}\lambda_{X}g_{XB_s\pi}}{(m_u+m_d)(m_b+m_s)} \frac{1}{\left(m_{X}^2-p^{\prime2}\right)\left(m_{B_s}^2-p^2\right)\left(m_{\pi}^2-q^2\right)}   \nonumber\\
&&+ \frac{C_{X\pi}}{\left(m_{X}^2-p^{\prime2}\right)\left(m_{B_s}^2-p^2\right)} + \frac{C_{XB_s}}{\left(m_{X}^2-p^{\prime2}\right)\left(m_{\pi}^2-q^2\right)} +\cdots\, .
\end{eqnarray}

We  set $p^{\prime2}=p^2$ and take  the double Borel transform with respect to the variable   $P^2=-p^2 $ and $Q^2=-q^2$ respectively to obtain the  QCD sum rule at the left side (LS),
\begin{eqnarray}
{\rm LS}&=&\frac{f_{\pi}m_{\pi}^2f_{B_s}m^2_{B_s}\lambda_{X}g_{XB_s\pi}}{(m_u+m_d)(m_b+m_s)} \frac{1}{m_{X}^2-m_{B_s}^2}  \left\{ \exp\left(-\frac{m_{B_s}^2}{M_1^2} \right)-\exp\left(-\frac{m_{X}^2}{M_1^2} \right)\right\}\exp\left(-\frac{m_{\pi}^2}{M_2^2} \right) \nonumber\\
&&+ C_{XB_s}\exp\left(-\frac{m_{X}^2}{M_1^2} \right)\exp\left(-\frac{m_{\pi}^2}{M_2^2} \right) \, .
\end{eqnarray}
In  calculations,   we neglect  the dependencies of the  $C_{X\pi}$  and $C_{X B_s}$  on the variables $p^2,\,p^{\prime 2},\,q^2$ therefore the dependencies of the  $C_{X\pi}$  and $C_{X B_s}$  on the variables $M_1^2$ and $M_2^2$, take the $C_{X\pi}$  and $C_{X B_s}$  as free parameters, and choose the suitable values  to
eliminate the contaminations so as  to obtain the stable sum rules with the variations of
the Borel parameters \cite{Ioffe-84,Wang-4200}.

\begin{figure}
 \centering
  \includegraphics[totalheight=2cm,width=6cm]{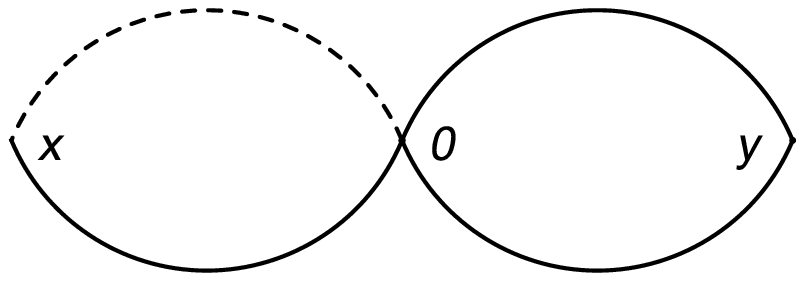}
 \includegraphics[totalheight=2cm,width=6cm]{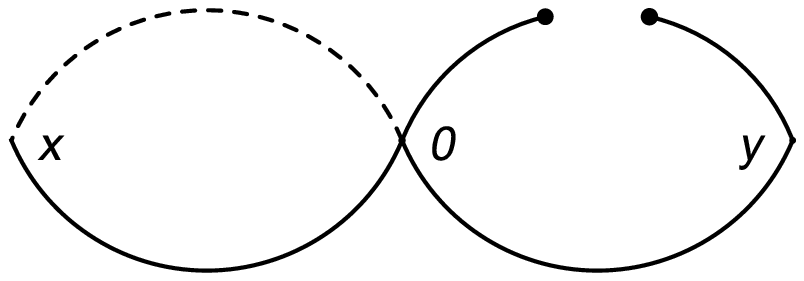}
\vglue+3mm
 \includegraphics[totalheight=2cm,width=6cm]{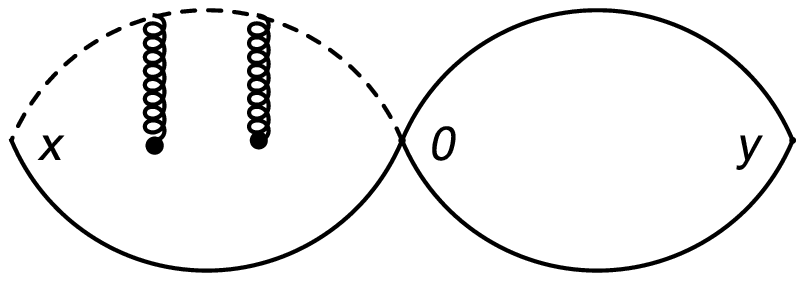}
 \includegraphics[totalheight=2cm,width=6cm]{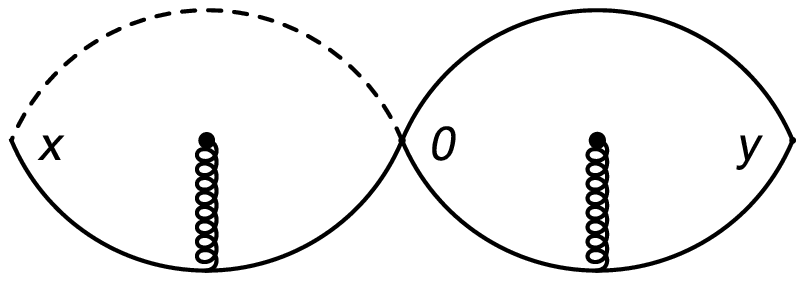}
 \vglue+3mm
 \includegraphics[totalheight=2cm,width=6cm]{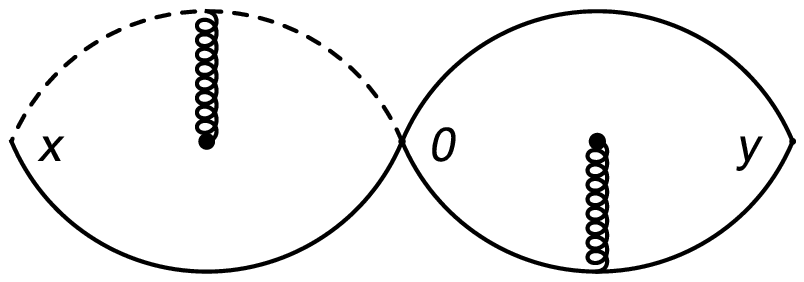}
  \includegraphics[totalheight=2cm,width=6cm]{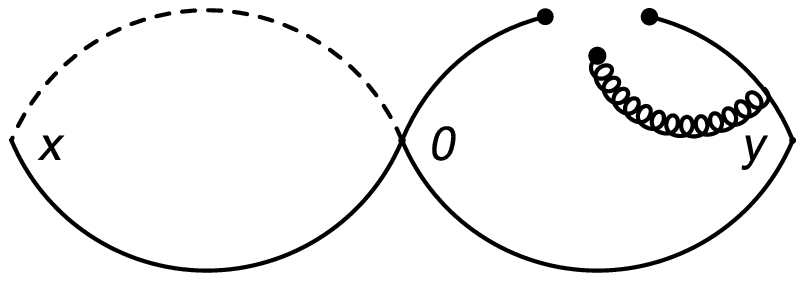}
  \vglue+3mm
 \includegraphics[totalheight=2cm,width=6cm]{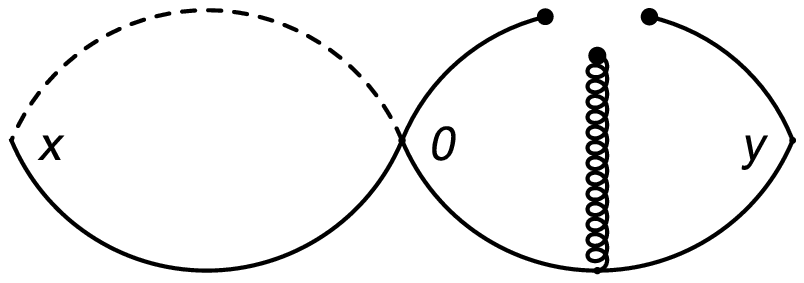}
 \includegraphics[totalheight=2cm,width=6cm]{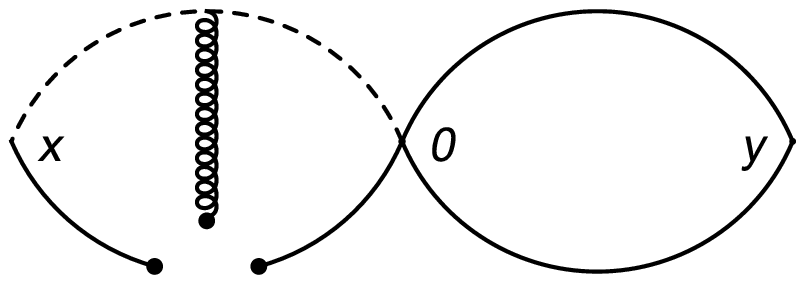}
\vglue+3mm
 \includegraphics[totalheight=2cm,width=6cm]{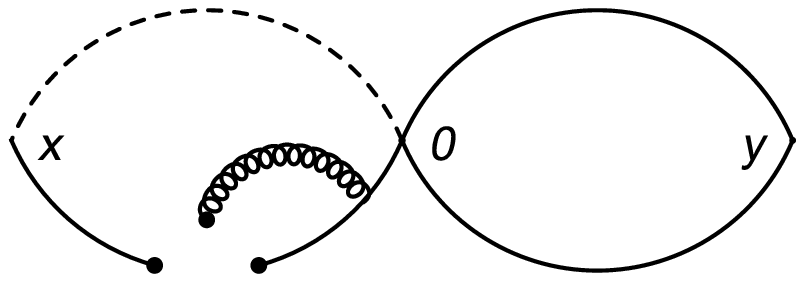}
  \includegraphics[totalheight=2cm,width=6cm]{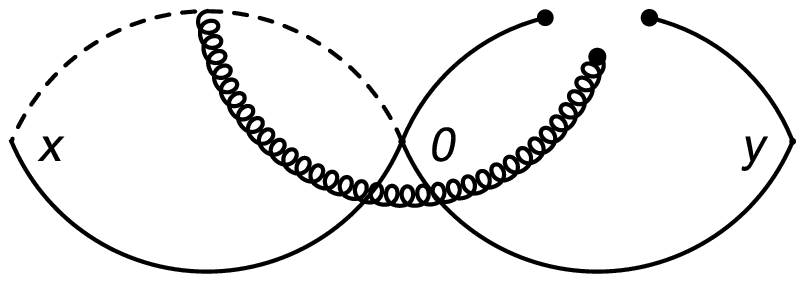}
  \vglue+3mm
 \includegraphics[totalheight=2cm,width=6cm]{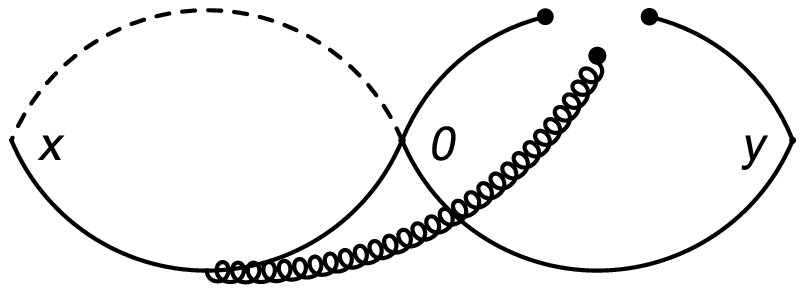}
  \includegraphics[totalheight=2cm,width=6cm]{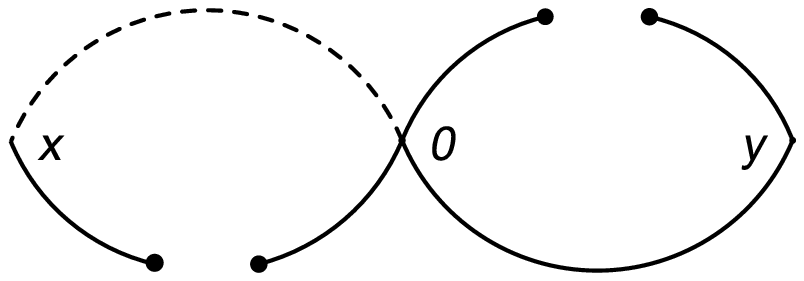}
 \caption{ The   Feynman diagrams  calculated in this article, where the solid lines and dashed lines denote the light quarks and heavy quarks, respectively, the waved lines denote the gluons. Other diagrams obtained  by interchanging of the light  quark lines are implied.  }
\end{figure}

Now we carry out the operator product expansion at the large Euclidean space-time region $-p^2 \to \infty$ and $-q^2 \to \infty$, take into account  the vacuum condensates up to dimension 6 and neglect the contribution of the three-gluon condensate, as the
three-gluon condensate  is    the vacuum expectation of the operator of the order  $\mathcal{O}( \alpha_s^{3/2})$.   In other words, we calculate the Feynman diagrams shown in Fig.1. For example, the first diagram is calculated in the following ways,
\begin{eqnarray}
\Pi(p,q)&=&-\frac{6}{(2\pi)^8}\int d^4k d^4l \frac{{\rm Tr} \left\{ \gamma_5 (\!\not\!{k}+m_s)\gamma_5 (\!\not\!{k}+\!\not\!{p}+m_b)\gamma_5(\!\not\!{l}+\!\not\!{q})\gamma_5 \!\not\!{l}\right\}}{k^2\left[(k+p)^2-m_b^2 \right]^2(l+q)^2 l^2} \nonumber\\
&=&-\frac{6}{(2\pi)^8} \frac{(-2\pi i)^2}{2\pi i} \int_{m_b^2}^\infty ds \frac{1}{s-p^2}\int d^4k\, \delta \left[ k^2\right] \delta \left[ (k+p)^2-m_b^2\right]
\frac{(-2\pi i)^2}{2\pi i} \int_{0}^\infty du \frac{1}{u-q^2}\nonumber\\
&&\int d^4l\, \delta \left[ l^2\right] \delta \left[ (l+q)^2\right]{\rm Tr} \left\{ \gamma_5 (\!\not\!{k}+m_s)\gamma_5 (\!\not\!{k}+\!\not\!{p}+m_b)\gamma_5(\!\not\!{l}+\!\not\!{q})\gamma_5 \!\not\!{l}\right\}\nonumber\\
&=& \frac{3}{128\pi^4}\int_{m_b^2}^\infty ds \frac{1}{s-p^2} \frac{(s-m_b^2)^2}{s}\int_{0}^\infty du \frac{u}{u-q^2}\nonumber\\
&& +\frac{3m_s m_b}{64\pi^4}\int_{m_b^2}^\infty ds \frac{1}{s-p^2} \frac{s-m_b^2}{s}\int_{0}^\infty du \frac{u}{u-q^2} \, .
\end{eqnarray}
The operator product expansion converges for large $-p^2$ and $-q^2$, it is odd to take the limit $q^2 \to 0$.

Then we set $p^{\prime2}=p^2$,  take the quark-hadron  duality below the continuum thresholds,  and perform   the double Borel transform with respect to the variables    $P^2=-p^2 $ and $Q^2=-q^2$ respectively to obtain the perturbative  term,
\begin{eqnarray}
B_{M_1^2,M_2^2}\Pi(p,q)&=&\frac{3}{128\pi^2} \int_{m_b^2}^{s_0}ds \int_0^{u_0}du \frac{(s-m_b^2)^2}{s}u\exp\left( -\frac{s}{M_1^2}-\frac{u}{M_2^2}\right)\nonumber\\
&&+\frac{3m_sm_b}{64\pi^2} \int_{m_b^2}^{s_0}ds \int_0^{u_0}du \frac{s-m_b^2}{s}u\exp\left( -\frac{s}{M_1^2}-\frac{u}{M_2^2}\right)\, ,
\end{eqnarray}
where the $s_0$ and $u_0$ are the continuum threshold parameters for the $X(5568)$ and $\pi$, respectively.

Other Feynman diagrams are calculated in  analogous ways, finally we obtain the QCD sum rules at the right side (RS),
\begin{eqnarray}
{\rm RS}&=&\frac{3}{128\pi^2} \int_{m_b^2}^{s_0}ds \int_0^{u_0}du \frac{(s-m_b^2)^2}{s}u\exp\left( -\frac{s}{M_1^2}-\frac{u}{M_2^2}\right)\nonumber\\
&&+\frac{3m_sm_b}{64\pi^2} \int_{m_b^2}^{s_0}ds \int_0^{u_0}du \frac{s-m_b^2}{s}u\exp\left( -\frac{s}{M_1^2}-\frac{u}{M_2^2}\right)\nonumber\\
&&+\frac{1}{192\pi^2 }\langle \frac{\alpha_sGG}{\pi}\rangle\int_{m_b^2}^{s_0}ds \int_{0}^{u_0}du \left(2-\frac{m_b^2}{s} \right)\exp\left( -\frac{s}{M_1^2}-\frac{u}{M_2^2}\right)\nonumber\\
&&-\frac{m_b\langle\bar{s}s\rangle}{16\pi^2}\int_0^{u_0}du u\exp\left( -\frac{m_b^2}{M_1^2}-\frac{u}{M_2^2}\right)\nonumber\\
&&-\frac{m_s\langle\bar{s}s\rangle}{32\pi^2}\left( 1+\frac{m_b^2}{M_1^2}\right)\int_0^{u_0}du u\exp\left( -\frac{m_b^2}{M_1^2}-\frac{u}{M_2^2}\right)\nonumber\\
&&+\frac{1}{192\pi^2 }\langle \frac{\alpha_sGG}{\pi}\rangle\int_0^{u_0}du u\exp\left( -\frac{m_b^2}{M_1^2}-\frac{u}{M_2^2}\right)\nonumber\\
&&+\frac{1}{128\pi^2 }\langle \frac{\alpha_sGG}{\pi}\rangle\int_{m_b^2}^{s_0}ds \frac{(s-m_b^2)^2}{s}\exp\left( -\frac{s}{M_1^2}\right)\nonumber\\
&&-\frac{m_b\langle\bar{s}g_s\sigma Gs\rangle}{32\pi^2 } \int_0^{u_0}du \left(1+\frac{u}{M_1^2}-\frac{u m_b^2}{2M_1^4}-\frac{u m_s m_b^3}{6M_1^6} \right)\exp\left( -\frac{m_b^2}{M_1^2}-\frac{u}{M_2^2}\right) \, .
\end{eqnarray}
The terms $\langle \bar{q}q\rangle\langle \bar{s}s\rangle$ disappear after performing the double  Borel transform, the last Feynman diagram in Fig.1 have no contribution.

In Refs.\cite{Wang-4200,Zhu-4200}, the width of the $Z_c(4200)$ is studied with the three-point QCD sum rules by including both the connected and disconnected Feynman diagrams, which is contrary to Ref.\cite{Nielsen3900}, where only the connected Feynman diagrams are taken into account to study the width of the $Z_c(3900)$. In this article, the contributions come from the connected diagrams can be written as ${\rm RS}_{c}$,	
\begin{eqnarray}
{\rm RS}_c&=&\frac{1}{192\pi^2 }\langle \frac{\alpha_sGG}{\pi}\rangle\int_{m_b^2}^{s_0}ds \int_{0}^{u_0}du \left(2-\frac{m_b^2}{s} \right)\exp\left( -\frac{s}{M_1^2}-\frac{u}{M_2^2}\right)\nonumber\\
&&-\frac{m_b\langle\bar{s}g_s\sigma Gs\rangle}{32\pi^2 } \int_0^{u_0}du \exp\left( -\frac{m_b^2}{M_1^2}-\frac{u}{M_2^2}\right) \, ,
\end{eqnarray}
which is too small to account for the experimental data \cite{X5568-exp}.

Finally, we obtain the QCD sum rule,
\begin{eqnarray}
{\rm  LS} &=&{\rm RS} \, .
\end{eqnarray}

There appear some energy scale dependence  at the hadron side (or LS) of the QCD sum rule according to the factors  $m_u+m_d$ and $m_b+m_s$, we can eliminate the energy scale dependence by using the currents $\widehat{J}_{B_s}(x)$ and $\widehat{J}_{\pi}(y)$,
\begin{eqnarray}
\widehat{J}_{B_s}(x)&=&\left(m_b+m_s \right)\bar{s}(x)i\gamma_5 b(x) \, ,\nonumber \\
\widehat{J}_{\pi}(y)&=&\left(m_u+m_d \right)\bar{u}(y)i\gamma_5 d(y) \, ,
\end{eqnarray}
then
\begin{eqnarray}
\langle0|\widehat{J}_{B_s}(0)|B_s(p)\rangle&=&f_{B_s}m_{B_s}^2 \,\, , \nonumber \\
\langle0|\widehat{J}_{\pi}(0)|\pi(q)\rangle&=&f_{\pi}m_{\pi}^2 \,\, ,
\end{eqnarray}
and
\begin{eqnarray}
C_{X\pi} &\to&C_{X\pi}\left(m_b+m_s \right)\left(m_u+m_d \right)\, \, ,\nonumber \\
C_{X B_s} &\to&  C_{X B_s} \left(m_b+m_s \right)\left(m_u+m_d \right)\, \, ,
\end{eqnarray}
the resulting QCD sum rule at the right side also acquires  a factor $\left(m_b+m_s \right)\left(m_u+m_d \right)$, a equivalent QCD sum rule is obtained,
the predicted hadronic coupling constant  $g_{XB_s\pi}$  is not changed.

We can also study the strong decay $X(5568)\to B_s^0\pi^{+}$  with the three-point correlation function
$\Pi_{\mu\nu}(p,q)$,
\begin{eqnarray}
\Pi_{\mu\nu}(p,q)&=&i^2\int d^4xd^4y e^{ip\cdot x}e^{iq\cdot y}\langle 0|T\left\{\eta_{\mu}^{\bar{s} b}(x)\eta_{\nu}^{\bar{u} d}(y)J_{X}(0)\right\}|0\rangle\, ,
\end{eqnarray}
where the currents
\begin{eqnarray}
\eta_{\mu}^{\bar{s} b}(x)&=&\bar{s}(x)\gamma_\mu\gamma_5 b(x) \, ,\nonumber \\
\eta_{\nu}^{\bar{u} d}(y)&=&\bar{u}(y)\gamma_\nu \gamma_5 d(y) \, ,
\end{eqnarray}
interpolate the mesons $B_s$ and $\pi$, respectively.
At the hadron side, we insert  a complete set of intermediate hadronic states with
the same quantum numbers as the current operators
    $\eta_{\mu}^{\bar{s} b}(x)$ and  $\eta_{\nu}^{\bar{u} d}(y)$  into the three-point
correlation function $\Pi_{\mu\nu}(p,q)$ and  isolate the ground state contributions to obtain the following result,
\begin{eqnarray}
\Pi_{\mu\nu}(p,q)&=&  \frac{f_{B_s}f_{\pi}\lambda_X g_{XB_s\pi}}{\left(m_{X}^2-p^{\prime2}\right)\left(m_{B_s}^2-p^2\right)\left(m_{\pi}^2-q^2\right)} \, \left( -p_\mu q_\nu\right)   \nonumber\\
&&+\frac{f_{B_{s1}}m_{B_{s1}}f_{\pi}\lambda_X g_{XB_{s1}\pi}}{\left(m_{X}^2-p^{\prime2}\right)\left(m_{B_{s1}}^2-p^2\right)\left(m_{\pi}^2-q^2\right)}
 \left( -q_\mu q_\nu +\frac{p \cdot q}{p^2} p_\mu q_\nu \right) \nonumber\\
&&+\frac{f_{B_{s1}}m_{B_{s1}}f_{a_1}m_{a_1}\lambda_X g_{XB_{s1} a_1}}{\left(m_{X}^2-p^{\prime2}\right)\left(m_{B_{s1}}^2-p^2\right)\left(m_{a_1}^2-q^2\right)}
 \left( g_{\mu\nu}  -\frac{1}{p^2} p_\mu p_\nu -\frac{1}{q^2} q_\mu q_\nu+\frac{p \cdot q}{p^2 q^2} p_\mu q_\nu\right) +\cdots \, ,\nonumber\\
\end{eqnarray}
where $p^\prime=p+q$, the $f_{B_{s1}}$, $f_{B_{s}}$, $f_{a_1}$ and $f_{\pi}$   are the decay constants of the mesons $B_{s1}(5830)$, $B_s$,  $a_1(1260)$ and $\pi$, respectively, the $g_{X B_{s1}\pi}$ and $g_{X B_{s1} a_1}$ are the hadronic coupling constants.

In the following, we write down the definitions,
\begin{eqnarray}
\langle0|\eta_{\mu}^{\bar{s} b}(0)|B_s(p)\rangle&=&i f_{B_s}p_\mu \,\, , \nonumber \\
\langle0|\eta_{\nu}^{\bar{u} d}(0)|\pi(q)\rangle&=&if_{\pi}q_\nu \,\, ,  \nonumber \\
\langle0|\eta_{\mu}^{\bar{s} b}(0)|B_{s1}(p)\rangle&=& f_{B_{s1}}m_{B_{s1}}\varepsilon_\mu \,\, , \nonumber \\
\langle0|\eta_{\nu}^{\bar{u} d}(0)|a_1(q)\rangle&=&f_{a_1}m_{a_1} \epsilon_\nu \,\, ,  \\
\langle B_{s1}(p)\pi(q)|X(p^{\prime})\rangle&=&\varepsilon^* \cdot q\, g_{X B_{s1}\pi} \,\, , \nonumber  \\
\langle B_{s1}(p)a_1(q)|X(p^{\prime})\rangle&=&i\varepsilon^* \cdot \epsilon^* \,g_{X B_{s1} a_1} \,\, ,
\end{eqnarray}
where the $\varepsilon_\mu$ and $\epsilon_\nu$ are  polarization vectors of the axialvector mesons $B_{s1}(5830)$ and  $a_1(1260)$, respectively.
From the values $m_X =\left(
5567.8 \pm 2.9 { }^{+0.9}_{-1.9}\right)\,\rm{MeV} $ \cite{X5568-exp}, $m_{B_{s1}}=\left(5828.40\pm 0.04\pm 0.41\right)\,\rm{MeV}$,  $m_{B_{s}}=\left(5366.7 \pm 0.4\right)\,\rm{MeV}$ \cite{PDG}, we can obtain $m_{B_{s1}}-m_{B_{s}}\approx462\,\rm{MeV}$ and $m_{B_{s1}}-m_{X}\approx 261\,\rm{MeV}$. If we take the interpolating currents
$\eta_{\mu}^{\bar{s} b}(x)$ and  $\eta_{\nu}^{\bar{u} d}(y)$, there are contaminations from the  axialvector mesons $B_{s1}(5830)$ and  $a_1(1260)$. We should multiply both sides of Eq.(19) by $p^\mu q^\nu$ to eliminate the contaminations of the axialvector mesons $B_{s1}(5830)$ and  $a_1(1260)$,
\begin{eqnarray}
p^\mu q^\nu \Pi_{\mu\nu}(p,q)&=&  \frac{f_{B_s}f_{\pi}\lambda_X g_{XB_s\pi}}{\left(m_{X}^2-p^{\prime2}\right)\left(m_{B_s}^2-p^2\right)\left(m_{\pi}^2-q^2\right)} \,\left(- p^2 q^2 \right)    +\cdots \, ,
\end{eqnarray}
which corresponds  to taking  the pseudoscalar currents $\widehat{J}_{B_s}(x)$ and $\widehat{J}_{\pi}(y)$ according to the following identities,
\begin{eqnarray}
\partial^\mu \eta_{\mu}^{\bar{s} b}(x)&=& \left(m_b+m_s\right) \bar{s}(x)i\gamma_5  b(x) =\widehat{J}_{B_s}(x)\, ,\nonumber \\
\partial^\nu\eta_{\nu}^{\bar{u} d}(y)&=&\left(m_u+m_d\right)\bar{u}(y)i \gamma_5 d(y)=\widehat{J}_{\pi}(y) \, .
\end{eqnarray}
The axialvector currents $\eta_{\mu}^{\bar{s} b}(x)$ and  $\eta_{\nu}^{\bar{u} d}(y)$ can also be chosen to study the strong decay $X(5568)\to B_s^0\pi^{+}$.

We also expect to study the strong decay $X(5568)\to B_s^0\pi^{+}$  with the light-cone QCD sum rules using the two-point correlation function
$\overline{\Pi}(p,q)$,
\begin{eqnarray}
\overline{\Pi}(p,q)&=&i\int d^4x e^{ip\cdot x}\langle \pi(q)|T\left\{J_{B_s}(x)J_{X}(0)\right\}|0\rangle\, ,
\end{eqnarray}
where  the $\langle \pi(q)|$ is an external $\pi$ state.

At the QCD side, we  obtain the following result after performing the wick's contraction,
\begin{eqnarray}
\overline{\Pi}(p,q)&=& i\int d^4x e^{ip\cdot x}\langle \pi(q)| \epsilon^{ijk}\epsilon^{imn} u^T_j(0) C\gamma_5 S_s^{kl}(-x) i\gamma_5 S_b^{ln}(x)\gamma_5C \bar{d}^T_m(0)|0\rangle \, ,
\end{eqnarray}
where   the $S_s^{kl}(-x)$     and $S_b^{ln}(x)$ are the full  $s$ and  $b$ quark propagators, respectively. The $u$ and $\bar{d}$ quarks stay at the same point $x=0$, the light-cone distribution amplitudes of the $\pi$ meson are almost useless, the integrals over the $\pi$ meson's 
light-cone distribution amplitudes reduce to overall normalization factors. In the light-cone QCD sum rules, such a situation is possible only in
the soft pion limit $q \to 0$, and the light-cone expansion
reduces to the short-distance  expansion \cite{Light-cone}. In Ref.\cite{AziziX5568-decay},  Agaev, Azizi and Sundu take the soft pion limit $q \to 0$, and choose the $C\gamma_\mu \otimes \gamma^\mu C$ type current to interpolate the $X(5568)$, and use the light-cone QCD sum rules to study the  strong decay $X(5568)\to B_s^0\pi^{+}$. The light-cone QCD sum rules are reasonable only in the soft pion approximation.

\section{Numerical results and discussions}
The hadronic parameters are taken as $m_{X}=5.5678\,\rm{GeV}$ \cite{X5568-exp}, $\lambda_{X}=6.7\times 10^{-3}\,\rm{GeV}^5$, $\sqrt{s_{0}}=(6.1\pm0.1)\,\rm{GeV}$  \cite{WangX5568}, $m_{\pi}=0.13957\,\rm{GeV}$,  $m_{B_s}=5.3667\,\rm{GeV}$ \cite{PDG},
  $f_{\pi}=0.130\,\rm{GeV}$, $\sqrt{u_{0}}=(0.85\pm0.05)\,\rm{GeV}$  \cite{Colangelo-Review},
$f_{B_s}=0.231 \,\rm{GeV}$  \cite{WangDCon},
 $f_{\pi}m^2_{\pi}/(m_u+m_d)=-2\langle \bar{q}q\rangle/f_{\pi}$ from the Gell-Mann-Oakes-Renner relation, and
$M_2^2=(0.8-1.2)\,\rm{GeV}^2$ from the QCD sum rules  \cite{Colangelo-Review}.
At the QCD side,  the  vacuum condensates are taken to be standard values,
$\langle\bar{q}q \rangle=-(0.24\pm 0.01\, \rm{GeV})^3$,  $\langle\bar{s}s \rangle=(0.8\pm0.1)\langle\bar{q}q \rangle$,
$\langle\bar{q}g_s\sigma G q \rangle=m_0^2\langle \bar{q}q \rangle$, $\langle\bar{s}g_s\sigma G s \rangle=m_0^2\langle \bar{s}s \rangle$,
$m_0^2=(0.8 \pm 0.1)\,\rm{GeV}^2$  and $\langle \frac{\alpha_s GG}{\pi}\rangle=(0.33\,\rm{GeV})^4$  at the energy scale  $\mu=1\, \rm{GeV}$
\cite{SVZ79,PRT85}.
The quark condensates and mixed quark condensates  evolve with the   renormalization group equation,
$\langle\bar{q}q \rangle(\mu)=\langle\bar{q}q \rangle(Q)\left[\frac{\alpha_{s}(Q)}{\alpha_{s}(\mu)}\right]^{\frac{4}{9}}$ and
 $\langle\bar{q}g_s \sigma Gq \rangle(\mu)=\langle\bar{q}g_s \sigma Gq \rangle(Q)\left[\frac{\alpha_{s}(Q)}{\alpha_{s}(\mu)}\right]^{\frac{2}{27}}$, where $q=u,d,s$.

In the article, we take the $\overline{MS}$ masses  $m_{b}(m_b)=(4.18\pm0.03)\,\rm{GeV}$ and $m_s(\mu=2\,\rm{GeV})=(0.095\pm0.005)\,\rm{GeV}$
 from the Particle Data Group \cite{PDG}, and take into account
the energy-scale dependence of  the $\overline{MS}$ masses from the renormalization group equation,
\begin{eqnarray}
m_b(\mu)&=&m_b(m_b)\left[\frac{\alpha_{s}(\mu)}{\alpha_{s}(m_b)}\right]^{\frac{12}{23}} \, ,\nonumber\\
m_s(\mu)&=&m_s({\rm 2GeV} )\left[\frac{\alpha_{s}(\mu)}{\alpha_{s}({\rm 2GeV})}\right]^{\frac{4}{9}} \, ,\nonumber\\
\alpha_s(\mu)&=&\frac{1}{b_0t}\left[1-\frac{b_1}{b_0^2}\frac{\log t}{t} +\frac{b_1^2(\log^2{t}-\log{t}-1)+b_0b_2}{b_0^4t^2}\right]\, ,
\end{eqnarray}
  where $t=\log \frac{\mu^2}{\Lambda^2}$, $b_0=\frac{33-2n_f}{12\pi}$, $b_1=\frac{153-19n_f}{24\pi^2}$, $b_2=\frac{2857-\frac{5033}{9}n_f+\frac{325}{27}n_f^2}{128\pi^3}$,  $\Lambda=213\,\rm{MeV}$, $296\,\rm{MeV}$  and  $339\,\rm{MeV}$ for the flavors  $n_f=5$, $4$ and $3$, respectively  \cite{PDG}.
 Furthermore, we set the $u$ and $d$ quark masses to be zero. In the heavy quark limit, the $b$-quark can be taken as a static potential well, and unchanged in the decay $X(5568)\to B_s^0 \pi^+$. In this article, we take the typical energy scale $\mu=m_b$.

The unknown parameter is chosen as $C_{XB_s}=-0.00059\,\rm{GeV}^8 $. There appears  a platform in the region $M_1^2=(4.5-5.5)\,\rm{GeV}^2$. Now we take into account the uncertainties of the input parameters and obtain the value of the hadronic coupling constant $g_{XB_s\pi}$, which is shown explicitly in Fig.2,
\begin{eqnarray}
g_{XB_s\pi}&=&\left(10.6\pm2.1 \right)   \,\rm{GeV} \, .
\end{eqnarray}

Now we obtain the partial decay width,
\begin{eqnarray}
\Gamma\left( X(5568) \to B_s^0 \pi^+\right) &=&\frac{g^2_{XB_s\pi}}{16\pi M_X^3} \sqrt{\left[m_X^2-(m_{B_s}+m_\pi)^2 \right]\left[m_X^2-(m_{B_s}-m_\pi)^2 \right]} \nonumber\\
&=&(20.5\pm8.1)\,\rm{MeV}\, .
\end{eqnarray}
The decays $X(5568) \to B^+\bar{K}^{0}$
  are kinematically forbidden, so the width $\Gamma_{X}$ can be saturated by the partial decay width $\Gamma\left( X(5568) \to B_s^0 \pi^+\right)$,  which is consistent with the experimental value $\Gamma_X = 21.9 \pm 6.4 {}^{+5.0}_{-2.5}\,\rm{MeV}$ from  the D0 collaboration \cite{X5568-exp}. The present work favors assigning the $X(5568)$ to be the scalar   diquark-antidiquark type tetraquark state.

\begin{figure}
 \centering
 \includegraphics[totalheight=6cm,width=8cm]{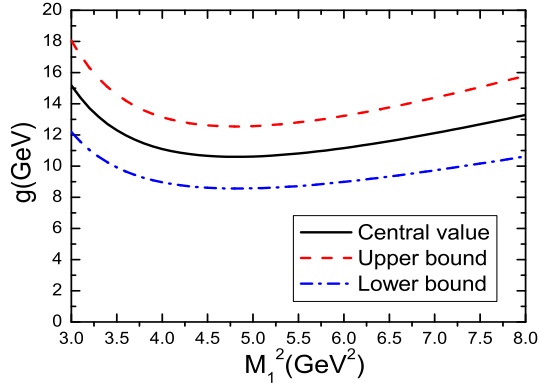}
         \caption{ The hadronic coupling constant  $g_{XB_s\pi}$  with variation of the Borel parameter $M_1^2$.  }
\end{figure}

In the following, we  perform Fierz re-arrangement to the current $J_X$ both in the color and Dirac-spinor  spaces to obtain the  result,
\begin{eqnarray}
J_X&=&\frac{1}{4}\left\{\,-\bar{b} s\,\bar{d} u+\bar{b}i\gamma_5 s\,\bar{d}i\gamma_5 u-\bar{b} \gamma^\mu s\,\bar{d}\gamma_\mu u
-\bar{b} \gamma^\mu\gamma_5 s\,\bar{d}\gamma_\mu\gamma_5 u+\frac{1}{2}\bar{b}\sigma_{\mu\nu} s\,\bar{d}\sigma^{\mu\nu} u\right. \nonumber\\
&&\left.+\bar{b} u\,\bar{d} s-\bar{b}i\gamma_5 u\,\bar{d}i\gamma_5 s+\bar{b} \gamma^\mu u\,\bar{d}\gamma_\mu c+
\bar{b} \gamma^\mu\gamma_5 u\,\bar{d}\gamma_\mu\gamma_5 s-\frac{1}{2}\bar{b}\sigma_{\mu\nu} u\,\bar{d}\sigma^{\mu\nu} s  \,\right\} \, ,
\end{eqnarray}
the components $\bar{b}i\gamma_5 s\,\bar{d}i\gamma_5 u$ and $\bar{b} \gamma^\mu\gamma_5 s\,\bar{d}\gamma_\mu\gamma_5 u$ couple potentially   to the meson pair   $B_s\pi^{+}$, while the components $\bar{b}i\gamma_5 u\,\bar{d}i\gamma_5 s$ and $\bar{b} \gamma^\mu\gamma_5 u\,\bar{d}\gamma_\mu\gamma_5 s$ couple potentially   to the meson pair   $B^+\bar{K}^0$. The strong decays
\begin{eqnarray}
X(5568) &\to& B_s\pi^{+} \, ,
\end{eqnarray}
are Okubo-Zweig-Iizuka  super-allowed, while the decays
\begin{eqnarray}
X(5568) &\to& B^+\bar{K}^{0} \, ,
\end{eqnarray}
  are kinematically forbidden, which is consistent with the observation of the D0 collaboration \cite{X5568-exp}. In previous works, we observed that the $C\gamma_5\otimes \gamma_5C$ type hidden-charm tetraquark states have slight smaller masses than that of the $C\gamma_\mu\otimes \gamma^\mu C$ type hidden-charm tetraquark states,  the predicted lowest masses  are
  $m_{C\gamma_5\otimes \gamma_5C}=\left(3.82^{+0.08}_{-0.08}\right)\,\rm{GeV}$ and
  $m_{C\gamma_\mu\otimes \gamma^\mu C}=\left(3.85^{+0.15}_{-0.09}\right)\,\rm{GeV}$ \cite{WangCmCm-C5C5}. We expect that  a $C\gamma_\mu\otimes \gamma^\mu C$ type current can also reproduce the experimental value $m_X =\left(
5567.8 \pm 2.9 { }^{+0.9}_{-1.9}\right)\,\rm{MeV} $ approximately \cite{ChenZhuX5568,AziziX5568-mass}.

Now we construct the current $\eta_{X}$ and  perform Fierz re-arrangement  both in the color and Dirac-spinor  spaces to obtain the following  result,
\begin{eqnarray}
\eta_{X}&=&\epsilon^{ijk}\epsilon^{imn}u^jC\gamma_\mu s^k \bar{d}^m\gamma^\mu C \bar{b}^n \, , \nonumber\\
 &=&  \bar{b}s\,\bar{d}u+\bar{b}i\gamma_5s\,\bar{d}i\gamma_5u+\frac{1}{2}\bar{b}\gamma_{\mu} s\,\bar{d}\gamma^{\mu}u-\frac{1}{2}\bar{b}\gamma_{\mu}\gamma_5 s\,\bar{d}\gamma^{\mu}\gamma_5u \nonumber\\
 &&+\bar{b}u\,\bar{d}s+\bar{b}i\gamma_5u\,\bar{d}i\gamma_5s+\frac{1}{2}\bar{b}\gamma_{\mu} u\,\bar{d}\gamma^{\mu}s-\frac{1}{2}\bar{b}\gamma_{\mu}\gamma_5 u\,\bar{d}\gamma^{\mu}\gamma_5s \, ,
\end{eqnarray}
the components $\bar{b}i\gamma_5 s\,\bar{d}i\gamma_5 u$ and $\bar{b} \gamma^\mu\gamma_5 s\,\bar{d}\gamma_\mu\gamma_5 u$ couple potentially   to the meson pair   $B_s\pi^{+}$, while the components $\bar{b}i\gamma_5 u\,\bar{d}i\gamma_5 s$ and $\bar{b} \gamma^\mu\gamma_5 u\,\bar{d}\gamma_\mu\gamma_5 s$ couple potentially   to the meson pair   $B^+\bar{K}^0$, which is analogous to the current $J_X$. It is also sensible  to assign the $X(5568)$ to be an axialvector-diquark-axialvector-antidiquark type tetraquark state or the  $X(5568)$ has some  axialvector-diquark-axialvector-antidiquark type tetraquark components.

The $C\otimes C$ type current $\widetilde{J}_{X}$ and $C\gamma_\mu\gamma_5 \otimes \gamma_5\gamma^\mu C $ type current $\widetilde{\eta}_{X}$ are expected to couple potentially to the scalar  tetraquark with much larger masses,
\begin{eqnarray}
\widetilde{J}_{X}&=&\epsilon^{ijk}\epsilon^{imn}u^jC s^k \bar{d}^m  C \bar{b}^n \, , \nonumber\\
 \widetilde{\eta}_{X}&=&\epsilon^{ijk}\epsilon^{imn}u^jC\gamma_\mu\gamma_5 s^k \bar{d}^m \gamma_5\gamma^\mu C \bar{b}^n \, ,
\end{eqnarray}
as the  favored configurations are the scalar diquarks ($C\gamma_5$-type) and axialvector diquarks ($C\gamma_\mu$-type)  from the QCD sum rules \cite{WangDiquark,WangLDiquark}.

\section{Conclusion}
In this article, we take the $X(5568)$  to be the scalar  diquark-antidiquark type tetraquark state,  study the hadronic coupling constant $g_{XB_s\pi}$ with  the three-point QCD sum rules, then calculate the partial decay width of the strong decay  $ X(5568) \to B_s^0 \pi^+$ and obtain the value $\Gamma_X=\left(20.5\pm8.1\right)\,\rm{MeV}$, which is   consistent with the experimental data $\Gamma_X = \left(21.9 \pm 6.4 {}^{+5.0}_{-2.5}\right)\,\rm{MeV}$ from the D0 collaboration. In calculation, we carry out the operator product expansion up to  the vacuum condensates of dimension-6, and take into account both the connected and disconnected Feynman diagrams.    The present prediction favors assigning  the $X(5568)$  to be the diquark-antidiquark type tetraquark state with $J^P=0^+$. However,
the quantum numbers  $J^P = 1^+$ cannot be excluded according to  decays  $X(5568) \to B_s^*\pi^+ \to B_s^0 \pi^+ \gamma$, where the low-energy photon is not
detected.

\section*{Acknowledgements}
This  work is supported by National Natural Science Foundation,
Grant Number 11375063, and Natural Science Foundation of Hebei province, Grant Number A2014502017.

\end{document}